\documentclass[aip,jap,reprint,superscriptaddress]{revtex4-2}

\usepackage[bookmarksnumbered,hyperfootnotes=false]{hyperref} 
\usepackage{graphicx} 
\usepackage{textcomp}
\usepackage{dcolumn}
\usepackage{bm}

\usepackage[utf8]{inputenc}
\usepackage[T1]{fontenc}
\usepackage{mathptmx}

\begin{document}



\title{What renders bulk metallic glass ductile/brittle? - new insight from the medium-range order}


\author{Farnaz A. Davani}
\affiliation{Institut f\"ur Materialphysik, Westf\"alische Wilhelms-Universit\"at M\"unster, Wilhelm-Klemm-Str. 10, 48149 M\"unster, Germany}

\author{Sven Hilke}
\affiliation{Institut f\"ur Materialphysik, Westf\"alische Wilhelms-Universit\"at M\"unster, Wilhelm-Klemm-Str. 10, 48149 M\"unster, Germany}

\author{Harald R\"osner}
\email[Corresponding author: ]{rosner@uni-muenster.de}
\affiliation{Institut f\"ur Materialphysik, Westf\"alische Wilhelms-Universit\"at M\"unster, Wilhelm-Klemm-Str. 10, 48149 M\"unster, Germany}

\author{David Geissler}
\affiliation{Leibniz IFW Dresden, Institute for Complex Materials (IKM), Helmholtzstr. 20, 01069 Dresden, Germany}

\author{Annett Gebert}
\affiliation{Leibniz IFW Dresden, Institute for Complex Materials (IKM), Helmholtzstr. 20, 01069 Dresden, Germany}

\author{Gerhard Wilde}
\affiliation{Institut f\"ur Materialphysik, Westf\"alische Wilhelms-Universit\"at M\"unster, Wilhelm-Klemm-Str. 10, 48149 M\"unster, Germany}


\date{\today}

\begin{abstract}
Understanding ductility or brittleness of monolithic bulk metallic glasses (BMGs) requires detailed knowledge of the amorphous structure. The medium range order (MRO) of ductile Pd$_{40}$Ni$_{40}$P$_{20}$ and brittle Zr$_{52.5}$Cu$_{17.9}$Ni$_{14.6}$Al$_{10}$Ti$_{5}$ (Vit105) was characterized prior to and after notched 3-point bending tests using variable-resolution fluctuation electron microscopy. Here we show the presence of a second larger MRO correlation length in the ductile material which is not present in the brittle material. A comparison with literature suggests that the larger correlation length accounts for larger shear transformation zones (STZs) which increase the heterogeneity. This enables an easier shear band formation and thus explains the difference in deformability.
\end{abstract}


\maketitle

\section{Introduction}
Bulk metallic glasses (BMGs) are of great interest as structural materials since they exhibit high strength and hardness. Most BMGs lack ductility, especially under tension where zero ductility prevails \cite{ashby2006metallic}. However, in recent years progress has been made in developing BMGs exhibiting respectable ductility during cold rolling, bending and compression tests. The key parameter for ductility seems to be the structural heterogeneity of the amorphous material, which can be selectively manipulated by imprinting, minor alloying, flash annealing, integrating crystalline phases to form composites or deformation-cycling in the so-called elastic regime \cite{schuh2007mechanical,eckert2007mechanical,scudino2011ductile,okulov2015flash,nollmann2016impact,scudino2018ductile,bian2019controlling,peng2019deformation,kosiba2019modulating}. In the absence of crystallinity, such structural heterogeneity may be comprehended as polyamorphism, i.e. the existence of two amorphous phases of the same composition separated by a first order transition \cite{sheng2007polyamorphism,zhu2015possible,pogatscher2018monotropic}; this concept also fits to the current understanding of nanoglasses \cite{gleiter2013nanoglasses}. 

Microstructurally, the structure-property relation of brittle/ductile behavior of MGs has been associated with the size (volume) of shear transformation zones (STZs), which carry the deformation. For brittle MGs, several investigations using rate-jump nanoindentation testing \cite{perriere2013nanoindentation,thurieau2015activation} or bending tests \cite{lei2019activation} showed relatively small STZ volumes of $\Omega_{STZ} = 0.2-0.8$\,nm$^3$. On the other hand, the STZ volumes of ductile BMGs were found to be larger, i.e. in the range of $\Omega_{STZ} = 6.0 - 6.5$\,nm$^3$ leading to the conclusion that larger STZ volumes are the reason for the good deformability \cite{pan2008experimental}. Interestingly, a recent molecular dynamics simulation \cite{hassani2019probing} suggested a correlation between medium-range order (MRO) and the size of STZs. 

To shed more light on the relation between MRO, STZs volumes and brittle/ductile behavior, the MRO of two representative materials for ductile/brittle, i.e. Pd$_{40}$Ni$_{40}$P$_{20}$ for ductile and Zr$_{52.5}$Cu$_{17.9}$Ni$_{14.6}$Al$_{10}$Ti$_{5}$ (Vit105) for brittle, was quantitatively investigated prior to and after deformation using variable-resolution fluctuation electron microscopy (VR-FEM) \cite{treacy1996variable,voyles2002fluctuation,treacy2005fluctuation,hwang2011variable,voyles2002fluctuatione}. The obtained MRO correlation lengths are compared with the size of STZs suggesting that the ductile behavior of BMGs correlates with the existence of additional, larger STZs or MRO size. Moreover, the structural heterogeneity in Pd$_{40}$Ni$_{40}$P$_{20}$ is discussed with respect to phosphorous and its covalency. 

\section{Methods}
To produce the Pd$_{40}$Ni$_{40}$P$_{20}$ alloy, ingots of Pd (99.5\,\%) and Ni$_2$P (99.9\,\%) were directly melted in a melt-spinning device under argon flow. For Zr$_{52.5}$Cu$_{17.9}$Ni$_{14.6}$Al$_{10}$Ti$_{5}$ (Vit105), the master alloy ingots/buttons for casting were prepared by arc-melting of pre-alloy buttons in a high purity Ti-gettered Ar atmosphere. A detailed description of these procedures is given by Geissler et al. \cite{geissler2019catastrophic}. Subsequently, 3-point bending tests of notched bars with dimensions of $17.75\times1.9\times0.88$\,mm (L$\times$W$\times$H) for Pd$_{40}$Ni$_{40}$P$_{20}$ and $20.5\times4.33\times2.24$\,mm for Vit105 were deformed as in-situ fracture and failure experiments (see Videos in the Supplementary Material).The aspect ratios of height H to width W are similar here, thus avoiding the known effects of sample size on the bend ductitility of metallic glass plates \cite{conner2003shear}.
 
Fluctuation electron microscopy (FEM) is a microscopic technique, which is sensitive to MRO using the four-body correlation of atom pairs (pair-pair correlation function) $g_4(|\vec{r}_1|, |\vec{r}_2|, |\vec{r}|, \theta)$ in disordered materials \cite{treacy1996variable}. A statistical analysis of the variance $V(k, R)$ from diffracted intensities of nanometer-sized volumes obtained by STEM microdiffraction was used here to extract this information. Sampling with different parallel coherent probe sizes, $R$, gives insight into the structural ordering length scale and provides a semi-quantitative measure of the MRO volume fraction (peak height or peak integral). This is called VR-FEM \cite{voyles2002fluctuation}. The normalized variance $V(k, R)$ of the spatially resolved diffracted intensity $I$ of a nanobeam diffraction pattern (NBDP) is thus a function of the scattering vector $\vec{k}$ and the coherent spatial resolution $R$:
\begin{equation}
V(|\vec{k}|,R) = \frac{\left\langle I^2 (\vec{k},R,\vec{r}) \right\rangle}{\left\langle I (\vec{k},R,\vec{r}) \right\rangle^2}-1
\label{EQ:NVAR}
\end{equation}
where $\left\langle\,\,\right\rangle$ indicates the averaging over different sample positions $\vec{r}$ or volumes respectively and R denotes the FWHM of the electron probe which defines the reciprocal space resolution \cite{treacy2005fluctuation}. We calculated the normalized variance profiles using a pixel by pixel analysis calculating the annular mean of variance image ($\Omega_{VImage} (k)$) \cite{daulton2010nanobeam}. 

VR-FEM and high-angle annular dark-field scanning transmission electron microscopy (HAADF-STEM) were performed at 300\,kV in a Thermo Fisher Scientific Themis 300 G3 transmission electron microscope (TEM). Electron-transparent lamellae of the deformed and the as-cast state were prepared by focused ion beam (FIB) \cite{hilke2019influence}. FEM analyses were performed on the diffracted intensities $I(\vec{k},R,\vec{r})$ of sets of individual NBDPs. For this purpose, $100-400$ NBDPs were acquired with parallel illumination using probe sizes between $0.8$ and $8.5$\,nm at FWHM. The \textmu Probe-STEM mode with the smallest C2 aperture (10\,\textmu m) was used. The electron probe size was subsequently varied by changing the semi-convergence angles ranging from 0.12\,mrad to 1.2\,mrad. Moreover, to minimize electron beam influences and to ensure equally high coherence, the nominal spot size 8 and a probe current of 15\,pA were used \cite{yi2010flexible,voyles2002fluctuatione}. The thicknesses of the FIB lamellae were determined from the low-loss part of electron energy loss (EEL) spectra using the log-ratio method so that equivalent conditions during the measurements were guaranteed \cite{malis1988eels}. It is worth noting that measurements within a sample were recorded under similar conditions for foil thickness, beam current and acquisition time in order to be sensitive to changes in peak heights and peak shapes of the normalized variance signal. To guarantee a good signal to noise ratio for the FEM analysis, the acquisition time was set to 4\,s for the individual NBDPs using the CCD camera (US 2000 Gatan) at binning 4 ($512\times512$ pixels). The camera lengths were set to 60\,mm for Pd$_{40}$Ni$_{40}$P$_{20}$ and 77\,mm for Vit105 to cover enough range in reciprocal space. All probe diameters used for the acquisition of the NBDPs were measured prior to the FEM experiments directly on the Ceta camera using Digital Micrograph plugins by D. Mitchell \cite{mitchell2005scripting}. 

\section{Models}
Gibson et al. \cite{gibson2000atom} introduced a method called pair-persistence analysis (PPA). It assumes that a metallic glass behaves like a supercooled liquid composed of small para-crystallites embedded in liquid-like matter as a simple model for the amorphous state. MRO is formed/present when such sub-structural units or motifs are correlated beyond the next-nearest neighbor distance. The four-atom positional correlation function $g_4(|\vec{r}_1|, |\vec{r}_2|, |\vec{r}|, \theta)$ yields the information of a characteristic correlation length $\Lambda$ of such MRO regions. This can be extracted from normalized variance profiles acquired with different probe sizes $R$ using the relation:
\begin{eqnarray}
\frac{Q^2}{V(|\vec{k}|,R)} = \underbrace{\left( \frac{1}{\Lambda^3 \cdot P(k)} \right)}_{intercept} + \underbrace{\left(\frac{4 \pi^2}{\Lambda \cdot P(k)}\right)}_{slope} \cdot Q^2 \nonumber\newline \\ \rightarrow \frac{Q^2}{V(|\vec{k}|,R)} = a + b \cdot Q^2 \rightarrow \Lambda(\text{nm}) = \frac{1}{2 \pi} \sqrt{\frac{b}{a}} \quad .
\label{EQ:PPA}
\end{eqnarray}
$\Lambda$ is calculated from the slope and intercept of the linear plot $Q^2/V$ against $Q^2$ with $Q = 0.61/R$.  However, this simplistic model assumes that the variance $V(k, Q)$ can be separated into a resolution and pair-persistence part ($V (k, Q) = R(Q) \cdot P(k)$). If more than one characteristic MRO correlation length exists, the results become either intercept- or slope-dominated leading to misinterpretation. Stratton and Voyles provide another approach which overcomes this problem:
\begin{equation}
V(|\vec{k}|,R) \approx a+b \cdot \frac{1}{R^2} \quad .
\label{EQ:Stratton-Voyles}
\end{equation}
The plot of $V(k, R)$ versus $1/R^2$ can also be divided into an intercept- and slope-dominated part \cite{stratton2008phenomenological,hwang2011variable}. However, it should be noted that this model requires sampling with many more probe sizes, including probe sizes $R > d$ (average MRO distance) in order to identify the a- and b-dominated parts properly. 

\begin{figure*}[htbp]
	\includegraphics[width=\textwidth]{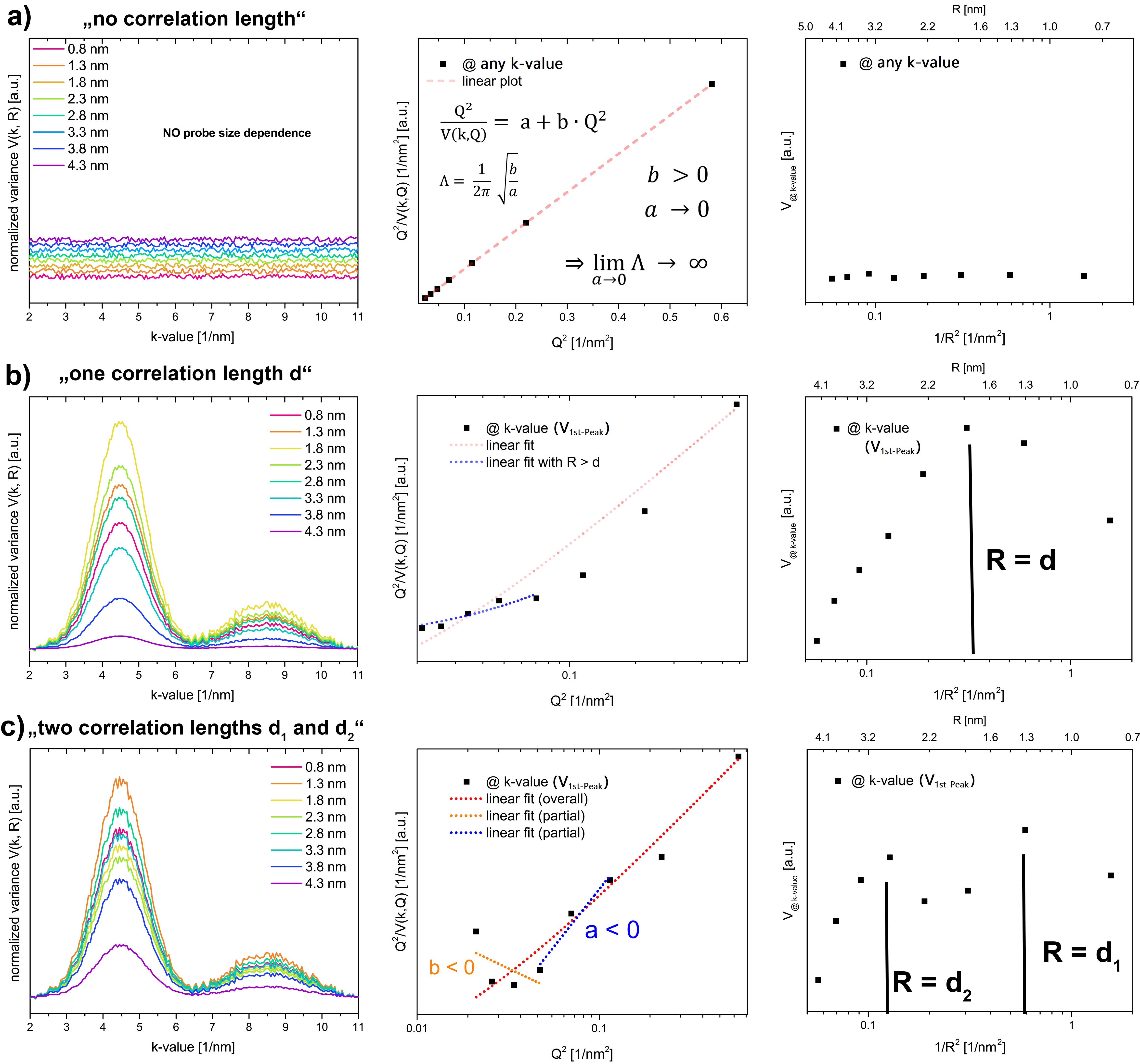}%
	\caption{Schematic assessment of the MRO using VR-FEM: (a) no MRO present, (b) MRO displaying one correlation length, (c) MRO with two correlation lengths. For each case, the left figure displays the normalized variance as a function of the k vector, the middle figure indicates the results obtained by the pair-persistence analysis (PPA) \cite{gibson2000atom} and the right figure summarizes the results obtained by the Stratton-Voyles method \cite{stratton2008phenomenological}.}
	\label{fig:FIG1}
\end{figure*}

Fig.~\ref{fig:FIG1} schematically illustrates these issues in detail: From typically more than 60 individual NBDPs the normalized variance is calculated. The plot of the normalized variance $V(k, R)$ against $k$ is structureless, if no MRO is present (Fig.~\ref{fig:FIG1}a), since there is no structural feature that is picked up by the probe. The absence of MRO results then in a variance (or standard deviation) that is equal to the mean. The PPA plot ($Q^2/(V(k, Q)$ against $Q^2$, see Fig. \ref{fig:FIG1}a-middle) reveals a very good linear fit to the data with an intercept going through the origin and thus yields an ``infinite'' correlation length $\Lambda$ according to Eq.~\ref{EQ:PPA}. The corresponding Stratton-Voyles plot ($V(k, R)$ versus $1/R^2$, see Fig.~\ref{fig:FIG1}a-right) is subsequently flat. On the other hand, if MRO is present and one correlation length between the structural motifs exists (Fig.~\ref{fig:FIG1}b-left), the plot of the normalized variance $V(k, R)$ against $k$ exhibits peaks. The probe size $R$ closest to the average MRO cluster distance results in larger peak to background heights of $V(k)$ reflecting that most clusters feature this correlation distance $d$. In Fig.~\ref{fig:FIG1}b this corresponds to $R = 1.8$\,nm. The corresponding PPA in Fig.~\ref{fig:FIG1}b-middle shows that different linear fits can be obtained when either smaller or larger probe sizes are used. The obtained correlation length $\Lambda$ is thus ambiguous. The Stratton-Voyles (Fig.~\ref{fig:FIG1}b-right) plot shows an increase with increasing probe size until the prevailing MRO distance $d$ is reached and then an abrupt decrease of the variance for larger probe sizes $R$. Moreover, the Stratton-Voyles plot illustrates the necessity of using larger and smaller probe sizes than the average MRO distance $d$ for the measurements, since otherwise the maximum is missed leading to misinterpretation of the data. The case of having two distinct correlation lengths is displayed in Fig.~\ref{fig:FIG1}c. At first glance Fig.~\ref{fig:FIG1}c-left does not look so much different from Fig.~\ref{fig:FIG1}b-left, where only one correlation length is present. The PPA plot (Fig.~\ref{fig:FIG1}c-middle) is difficult to interpret. However, the Stratton-Voyles model (Fig.~\ref{fig:FIG1}c-right) exhibits a double-peak feature showing the presence of two different MRO distances $d$. Thus, the Stratton-Voyles approach is sensitive to structural heterogeneity to the extent that different MRO correlation lengths can be discriminated and quantified. 

\section{Results}

\begin{figure*}[htbp]
	\includegraphics[width=0.9\textwidth]{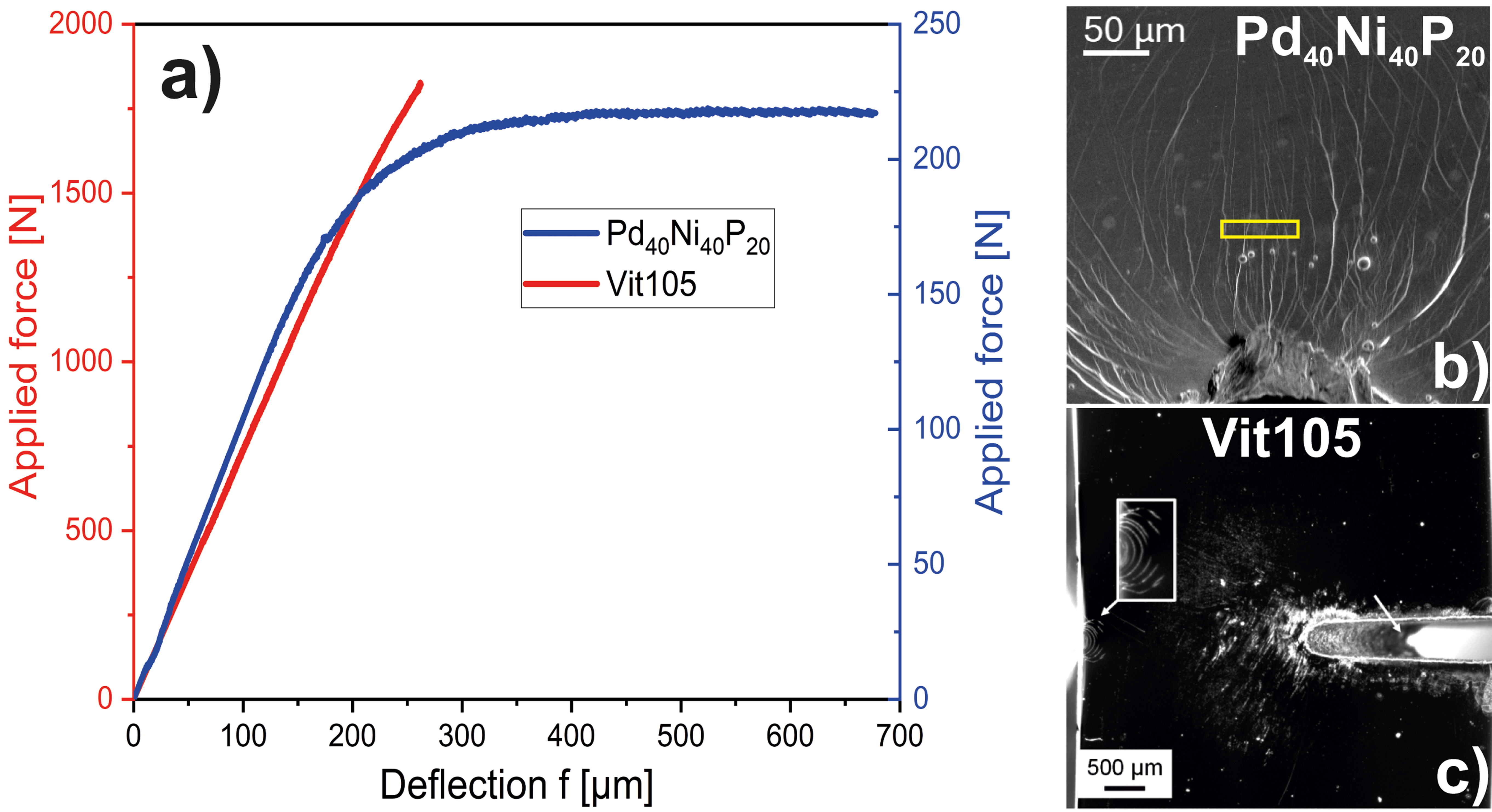}%
	\caption{(a) In-situ fracture and failure experiments of notched BMGs: ductile Pd$_{40}$Ni$_{40}$P$_{20}$ (blue curve) and brittle Vit105 (Zr$_{52.5}$Cu$_{17.9}$Ni$_{14.6}$Al$_{10}$Ti$_{5}$), red curve. (b) SEM micrograph of Pd$_{40}$Ni$_{40}$P$_{20}$ showing SBs around the notch area (tension zone). The yellow framed area indicates the region from which the FIB lamella was taken. (c) SEM micrograph of Vit105. The shear band region of the opposite side of the sample (compressive zone) is shown enlarged in the white frame. For more details see the Supplementary Material.}
	\label{fig:FIG2}
\end{figure*}

Detailed results of the 3-point bending tests of Pd$_{40}$Ni$_{40}$P$_{20}$ and Zr$_{52.5}$Cu$_{17.9}$Ni$_{14.6}$Al$_{10}$Ti$_{5}$ (Vit105) are shown as videos in the Supplementary Material. The corresponding force-deflection curves are shown in Fig.~\ref{fig:FIG2}. The Pd$_{40}$Ni$_{40}$P$_{20}$ alloy showed a continuous ductile behavior during the 3-point bending test (Fig.~\ref{fig:FIG2}a) \cite{nollmann2016impact}. However, the experiment was deliberately stopped before failure \cite{geissler2019catastrophic} (see Video of Pd$_{40}$Ni$_{40}$P$_{20}$ in the Supplementary Material). On the other hand, the Zr-based BMG (red curve in Fig.~\ref{fig:FIG2}a) failed shortly after the end of the elastic range. It is worth noting that shear bending was observed in the notched area of the tensile side as well as on the compressive side in both materials (for Pd$_{40}$Ni$_{40}$P$_{20}$ see Fig.~\ref{fig:FIG2}b and for Vit105 see Fig.~\ref{fig:FIG2}c). As a result shear steps were observed at the surface. 

\begin{figure}[htbp]
	\includegraphics[width=\linewidth]{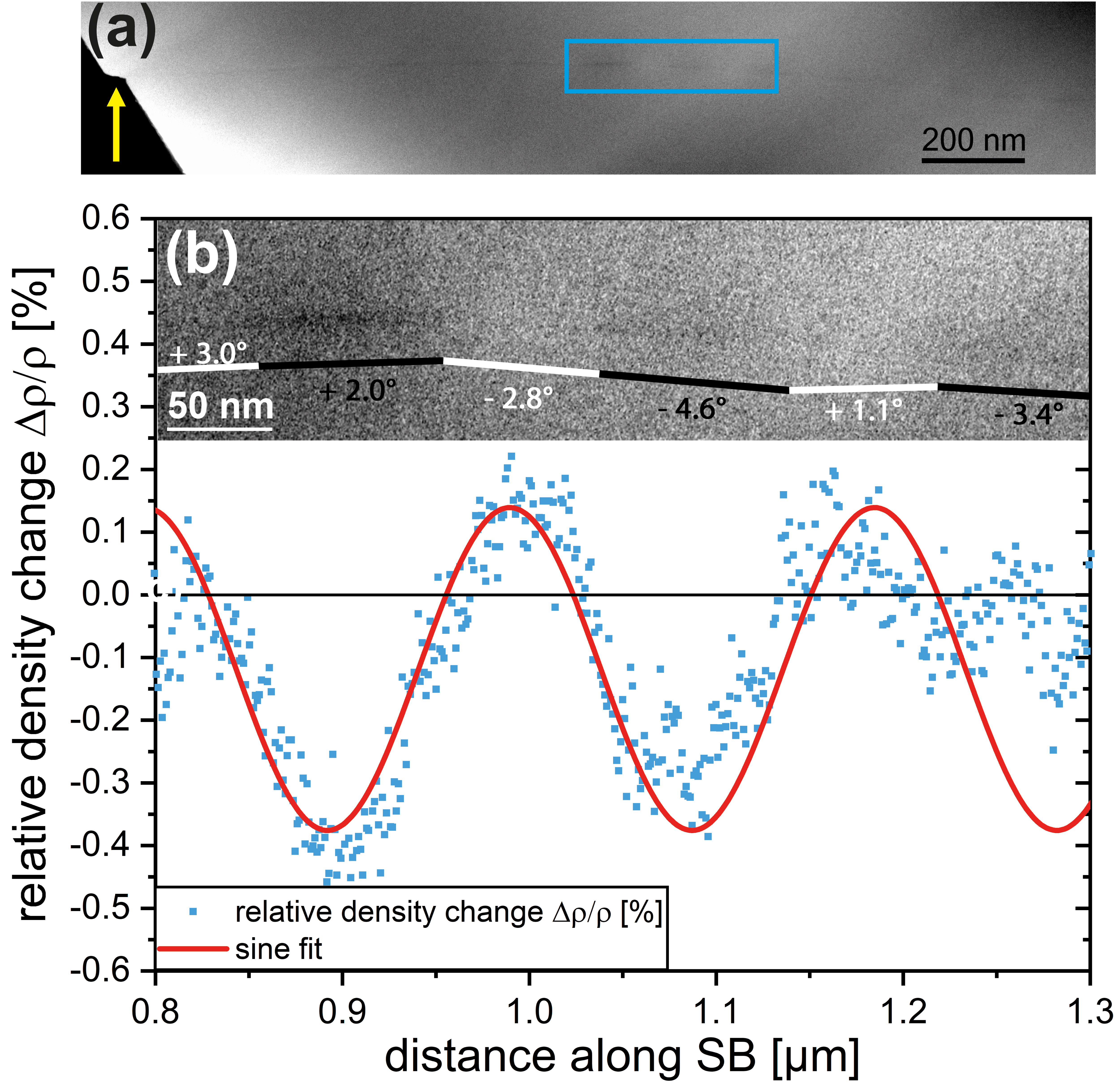}%
	\caption{(a) HAADF-STEM micrograph of a SB in Pd$_{40}$Ni$_{40}$P$_{20}$. Note the characteristic alternating dark and bright contrast changes along the SB. The yellow arrow indicates the shear step at the surface. (b) Enlargement of the blue marked area in (a). The relative density changes were calculated along the SB according to \cite{rosner2014density,schmidt2015quantitative,hieronymus2017shear,hilke2019influence}. Note the deflections of the SB.}
	\label{fig:FIG3}
\end{figure}

For both materials, comprehensive TEM studies were performed on shear bands that penetrated through the surfaces around the notched areas. The results for Vit105 can be found in \cite{hilke2019influence}. For Pd$_{40}$Ni$_{40}$P$_{20}$, the results are shown in Fig.~\ref{fig:FIG3}. The HAADF-STEM micrograph (Fig.~\ref{fig:FIG3}a) shows a representative SB present in the FIB lamella that can be followed from the surface (shear step) down to a depth of about 2\,\textmu m. The visibility of the SB varies over the whole distance. This is due to the bending of the SB which misaligns the SB to the electron beam so that the contrast between SB and matrix averages out along the projection. The part of the SB with the best visibility (blue marked area) was quantified with respect to its mass density \cite{rosner2014density,schmidt2015quantitative,hieronymus2017shear,hilke2019influence}. The results are shown in Fig.~\ref{fig:FIG3}b. Bright and dark contrast alternations along the shear band accompanied by small deflections between the SB segments are observed similar to SB studies on other MGs \cite{rosner2014density,schmidt2015quantitative,hieronymus2017shear,hilke2019influence}. A recent molecular dynamics (MD) simulation was able to connect the degree of deflections of the SB segments with the temporal evolution of the stress-strain curve \cite{hassani2019probing}. SBs originating from the early stages of the stress-strain curve showed larger deflection angles for the SB segments than those formed during the later stages of deformation. Applying this knowledge from the MD simulation to the SB observed in Fig.~\ref{fig:FIG3}a suggests that it is a ``young'' SB. This is further confirmed by the small shear step of about 50\,nm at the sample surface (indicated by the yellow arrow in Fig.~\ref{fig:FIG3}a). The maximum amplitude of the relative density change (see bottom of Fig.~\ref{fig:FIG3}b) seems to be slightly smaller than for Vit105 \cite{hilke2019influence}. However, the general appearance of the shear band (density alterations and deflections) does not reveal any significant difference that would immediately account for ductile or brittle behavior. Thus, the presence of shear bands alone cannot explain the difference in the mechanical behavior (ductile/brittle). In order to obtain further information on what actually defines ductile/brittle behavior, the amorphous structure of un-deformed and deformed states was investigated by VR-FEM. 

The full data sets of the normalized variance plots are displayed in the Supplementary Material. The sample thickness was calculated to be $(90\pm17)$\,nm for Pd$_{40}$Ni$_{40}$P$_{20}$ and $(52\pm11)$\,nm for Vit105 \cite{malis1988eels}. 

\begin{figure}[htbp]
	\includegraphics[width=\linewidth]{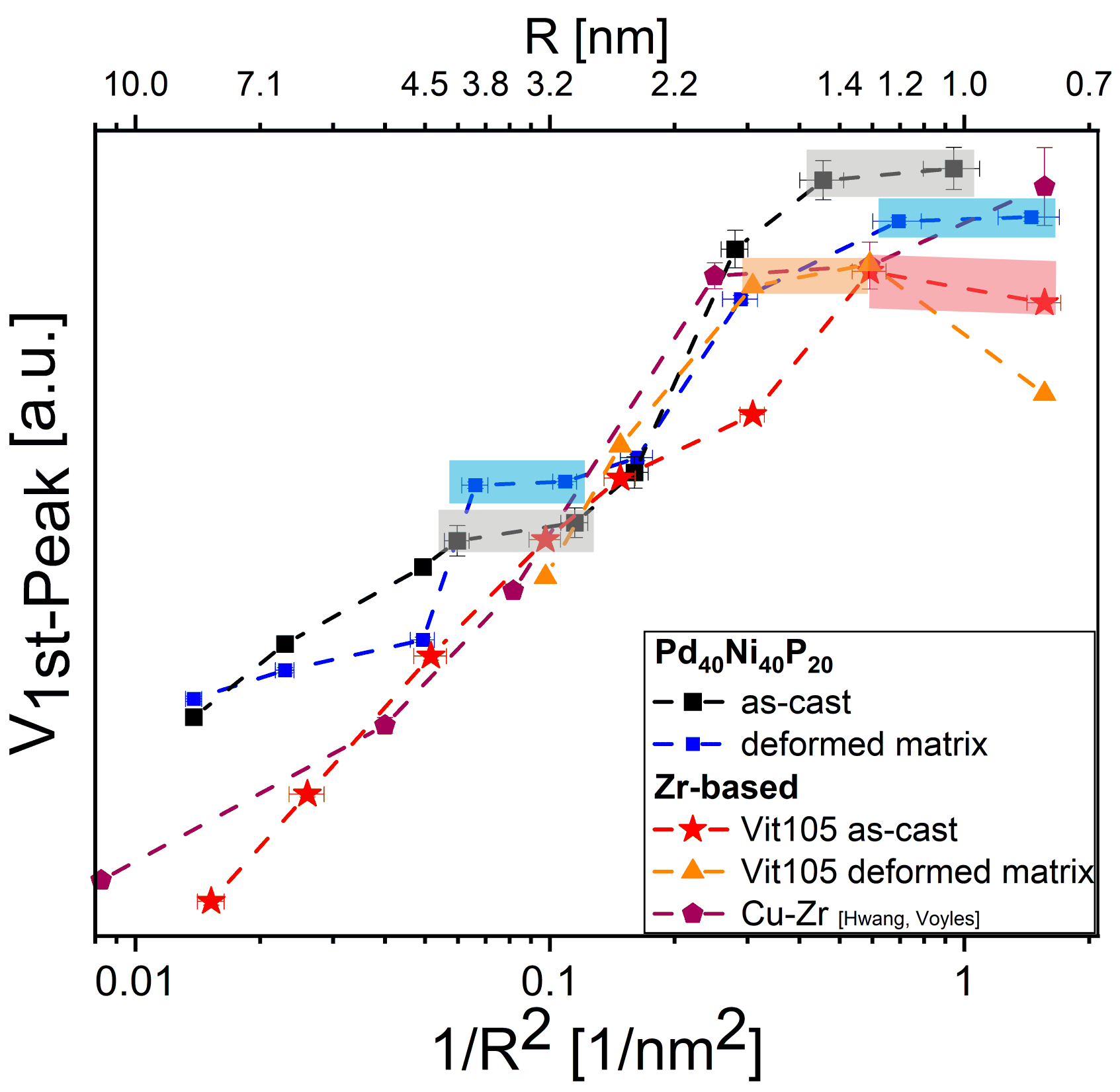}%
	\caption{FEM analysis of the amorphous structure showing $V(R)$ plotted against $1/R^2$ following Stratton-Voyles \cite{stratton2008phenomenological}. Note the difference: two plateaus for Pd$_{40}$Ni$_{40}$P$_{20}$ as indicated by the individual colored areas, but only one plateau for Vit105 and CuZr \cite{hwang2011variable}.}
	\label{fig:FIG4}
\end{figure}

The Stratton-Voles approach \cite{stratton2008phenomenological} was applied here since it allows inspection of the underlying amorphous structure relating to one or more prevailing MRO sizes. The initial state (as-cast state) as well as the deformed states of Pd$_{40}$Ni$_{40}$P$_{20}$ reveal two plateaus due to two MRO correlation lengths. For Vit105 (Fig.~\ref{fig:FIG4}), as well as the literature data for CuZr \cite{hwang2011variable}, only one MRO correlation length (one plateau in Fig.~\ref{fig:FIG4}) was observed. Using the observed MRO correlation lengths (see plateaus in Fig.~\ref{fig:FIG4}) to estimate corresponding volumes for the MRO regions (assuming here a spherical shape of the MRO with the correlation length d as diameter) yields the results summarized in Tab.~\ref{tab:TAB1}. The question arises whether this difference in structural heterogeneity in terms of the MRO correlation lengths distribution unambiguously discriminates whether a BMG is ductile or brittle? 

\section{Discussion}
In the following we discuss two main points: Firstly, the connection between structure (here MRO sizes), the STZ volumes and the mechanical behavior and secondly, the structural motifs in our BMGs with special respect to the role of the phosphorous and its covalency. 

A former study reports that the ductility of BMGs intrinsically correlates with their STZs volumes \cite{pan2008experimental}. This explains why an improved deformability/ductility of MGs was observed after imprinting, minor alloying, flash annealing, integrating crystalline phases to form composites or deformation-cycling in the so-called elastic regime \cite{eckert2007mechanical,scudino2011ductile,okulov2015flash,nollmann2016impact,scudino2018ductile,zhang2018dual,bian2019controlling,peng2019deformation,kosiba2019modulating,schawe2019existence,csopu2019structure} since all these procedures stimulate topological changes by shuffling atoms/clusters and thus increase the STZ volume. The results obtained from Fig.~\ref{fig:FIG4} (see Tab.~\ref{tab:TAB1}) are now compared with the STZ volumes from literature data. 

The estimated MRO volume of $\Omega_{STZ} = (0.52 \pm 0.21)$\,nm$^3$ for as-cast Vit105 shows an excellent agreement with the literature values of STZ volumes of various brittle MGs $\Omega_{STZ} = (0.2 - 0.8)$\,nm$^3$ (Tab.~\ref{tab:TAB1}) \cite{perriere2013nanoindentation,thurieau2015activation,lei2019activation}. On the other hand, at first glance the estimated volumes ($V_1$ or $V_2$) for as-cast Pd$_{40}$Ni$_{40}$P$_{20}$ do not agree with the reported literature value of about $\Omega_{STZ} = 6.5$\,nm$^3$ \cite{pan2008experimental}. However, it should be noted that the STZ volume $\Omega_{STZ}$ measured by rate-jump nanoindentation testing is an average value from a macroscopic sample that cannot distinguish microscopically between individual fractions of STZ volumes. Our measurements clearly reveal two MRO correlation lengths for the amorphous structure of Pd$_{40}$Ni$_{40}$P$_{20}$ and thus two contributions; i.e. MRO fractions $\Phi_1 (d_1, V_1)$ and $\Phi_2 (d_2, V_2)$. Hence, the difference between our microscopic results and the literature value from macroscopic measurements can be comprehended, if we consider both estimated MRO volumes in Tab.~\ref{tab:TAB1} to be fractions averaging to the literature value of $\Omega_{STZ}$(Pd$_{40}$Ni$_{40}$P$_{20}$)$ = 6.5$\,nm$^3$. As a result, the corresponding MRO fractions can be calculated using the following formulae (applying the lever rule for the fractions): 
\begin{eqnarray}
\Phi_1 \cdot V_1 + \Phi_2 \cdot V_2 = \Omega_{STZ} \nonumber\newline\\ \quad \text{with} \quad \Phi_1 + \Phi_2 =1 \Rightarrow \Phi_1 = \frac{\Omega_{STZ} - V_2}{V_1 - V_2} \,.
\label{EQ:STZ}
\end{eqnarray}
Using Eq.~\ref{EQ:STZ} with the two estimated MRO volumes ($V_1, V_2$) as input, the relative MRO volume fractions are calculated to be $\Phi_1 = 0.74$ and $\Phi_2 = 0.26$. As a result, roughly a quarter of the MRO is correlated over larger distances resembling larger STZ volumes. Thus, an increased heterogeneity enables an easier shear band formation promoting the ductility in Pd$_{40}$Ni$_{40}$P$_{20}$. 

\begin{table*}[htbp]
	\caption{Volumes estimated from the observed MRO correlation lengths using the mean value of the plateaus shown in Fig.~\ref{fig:FIG4}. Further information regarding the data evaluation is given in the Supplementary Material.}
	\vspace{5pt}
	\begin{ruledtabular}
		\begin{tabular}{c||c|c|c|c}
			material [state] & correlation length $d_1$ [nm] & volume $V_1$ [nm$^3$] & correlation length $d_2$ [nm] & volume $V_2$ [nm$^3$] \\  
			\hline \hline
			Vit105 [as-cast] & $(1.0\pm0.2)$ & $(0.52\pm0.21)$ & - & -\\ 
			Vit105 [deformed] & $(1.5\pm0.3)$ & $(1.77\pm0.71)$ & - & -\\ 
			\hline
			Pd$_{40}$Ni$_{40}$P$_{20}$ [as-cast] & $(1.2\pm0.2)$ & $(0.90\pm0.31)$ & $(3.50\pm0.60)$ & $(22.5\pm7.7)$\\ 
			Pd$_{40}$Ni$_{40}$P$_{20}$ [deformed] & $(1.0\pm0.2)$ & $(0.52\pm0.21)$ & $(3.45\pm0.50)$ & $(21.5\pm6.2)$\\ 
		\end{tabular} 
	\end{ruledtabular}
	\label{tab:TAB1}
\end{table*}

Now we discuss the significance of structural motifs in MGs as elements to form MRO networks \cite{lee2011networked}. The second correlation length in Pd$_{40}$Ni$_{40}$P$_{20}$ corresponds to another MRO network \cite{wang2018spatial}. The phosphorous in Pd$_{40}$Ni$_{40}$P$_{20}$ introduces the possibility of covalent bonding (metal-metalloid) through hybridization to form a specific class of clusters. Therefore, it is likely that the phosphorous leads to an increase in the overall variety of structural motifs in addition to the prevailing icosahedral motifs, which are present in the Zr-based glasses \cite{zhang2014composition}. Recent reports showed the existence of almost perfect icosahedral configurations around the Ni atoms for Pd$_{40}$Ni$_{40}$P$_{20}$, whereas the local configurations around Pd and P atoms resemble rather highly distorted icosahedral clusters \cite{hosokawa2019partial}. A molecular dynamics (MD) simulation came to a similar conclusion, finding the coexistence of P-centered polyhedra, that is tricapped trigonal prisms, and densely packed Ni-centered icosahedra leading to a topological order between these two clusters \cite{guan2012structural}. Furthermore, it was observed by another MD simulation combined with the recently developed motif extraction that for some motifs the population was strongly altered by by-passing the glass transition using an ultrafast cooling \cite{maldonis2019short,maldonis2019local}. It seems that for the metal-metalloid case, a bicapped square antiprism motif (in addition to the icosahedral motif \cite{maldonis2019short}) plays a significant role in the glass transition \cite{maldonis2019local}. The influence of covalent bonding seems to be the key parameter for the frustration of icosahedral structures \cite{maldonis2019local,zhan2017effect}. Thus, the observed increase in the variety of structural motifs as well as their population in the glassy state for Pd$_{40}$Ni$_{40}$P$_{20}$ enables the possibility of forming an enhanced heterogeneity \cite{wang2018spatial} over extended medium range affecting its plasticity, similar to observations in PdSi \cite{maldonis2019local,moitzi2020chemical} and AlSm \cite{Maldonis2019}. The structural heterogeneity may also be classified as a sort of polyamorphism; not, however, in the strict sense of a first order transformation between two phases but rather as distinguishable structural MRO networks.

\section{Conclusions}
The aim of this work was to elucidate a correlation between the plasticity of BMGs and their amorphous structure. For this purpose, two different types of BMGs, ductile Pd$_{40}$Ni$_{40}$P$_{20}$ and brittle Zr$_{52.5}$Cu$_{17.9}$Ni$_{14.6}$Al$_{10}$Ti$_{5}$ (Vit105), were investigated by TEM/FEM. While the nature of the shear bands did not reveal any significant differences to explain the completely different deformation behavior (brittle/ductile), VR-FEM analysis revealed distinct differences between the MRO structure, i.e. two correlation lengths in the case of Pd$_{40}$Ni$_{40}$P$_{20}$. A comparison of the smaller MRO correlation length with the size of STZs from literature showed an excellent agreement. The larger MRO correlation length corresponded to larger STZs, which increase the heterogeneity. This, we believe, enables an easier shear band formation promoting the ductility in Pd$_{40}$Ni$_{40}$P$_{20}$.

\section*{SUPPLEMENTARY MATERIAL}
See supplementary material for the complete data sets of the variable resolution fluctuation electron microscopy acquisition as well as the videos of the 3-point bending tests.

\section*{AUTHOR'S CONTRIBUTIONS}
All authors contributed equally to this work.

\section*{DATA AVAILABILITY}
The original data of this study are available from the corresponding author upon reasonable request. Moreover, the analyzed data that supports the findings of this study are available within the article and its supplementary material.

\begin{acknowledgments}
We gratefully acknowledge financial support by the DFG via SPP 1594 (Topological engineering of ultra-strong glasses, WI 1899/27-2 and GE 1106/11) and WI 1899/29-1 (Coupling of irreversible plastic rearrangements and heterogeneity of the local structure during deformation of metallic glasses, project number 325408982). The DFG is further acknowledged for funding our TEM equipment via the Major Research Instrumentation Program under INST 211/719-1 FUGG. 
\end{acknowledgments}

\section*{REFERENCES}
\bibliography{literature}

\end{document}